\documentclass[useAMS,usenatbib]{mn2e}

\usepackage{graphicx,color}

\title[The detailed chemical composition of Kepler-10]
{The detailed chemical composition of the terrestrial planet host Kepler-10\thanks{Based on observations obtained at the 3.6m Canada-France-Hawaii Telescope located at the Mauna Kea Observatory, US, the 9.2 m Hobby-Eberly Telescope located at the W.J. McDonald Observatory of the University of Texas at Austin, US, and the 6.5 m Magellan Clay Telescope located at the Las Campanas Observatory, Chile.}}

\author[F. Liu et al.]{F. Liu,$^{1}$\thanks{E-mail:fan.liu@anu.edu.au}
D. Yong,$^{1}$
M. Asplund,$^{1}$
I. Ram\'irez,$^{2}$
J. Mel\'endez,$^{3}$
B. Gustafsson,$^{4,5}$
\newauthor
L. M. Howes,$^{1,7}$
I. U. Roederer,$^{6}$
D. L. Lambert$^{2}$ and
T. Bensby$^{7}$\\
$^{1}$Research School of Astronomy and Astrophysics, Australian National University, Canberra, ACT 2611, Australia\\
$^{2}$McDonald Observatory and Department of Astronomy, University of Texas at Austin, 2515 Speedway, Austin, TX 78712-1205, USA\\
$^{3}$Departamento de Astronomia do IAG/USP, Universidade de Sao Paulo, Rua do Matao 1226, Sao Paulo 05508-900, SP, Brasil\\
$^{4}$Institution for Physics and Astronomy, Uppsala University, Box 515, SE-75120 Uppsala, Sweden\\
$^{5}$Nordita, Roslagstullsbacken 23, SE-10691 Stockholm, Sweden\\
$^{6}$Department of Astronomy, University of Michigan, 1085 South University Avenue, Ann Arbor, MI 48109, USA\\
$^{7}$Lund Observatory, Department of Astronomy and Theoretical physics, Lund University, Box 43, 22100 Lund, Sweden}

\begin{document}

\date{Accepted 2015 November 29. Received 2015 November 27; in original form 2015 April 16}

\pagerange{\pageref{firstpage}--\pageref{lastpage}} \pubyear{2015}

\maketitle

\label{firstpage}

\begin{abstract}
Chemical abundance studies of the Sun and solar twins have demonstrated that the solar composition of refractory elements is depleted when compared to volatile elements, which could be due to the formation of terrestrial planets. In order to further examine this scenario, we conducted a line-by-line differential chemical abundance analysis of the terrestrial planet host Kepler-10 and fourteen of its stellar twins. Stellar parameters and elemental abundances of Kepler-10 and its stellar twins were obtained with very high precision using a strictly differential analysis of high quality CFHT, HET and Magellan spectra. 
When compared to the majority of thick disc twins, Kepler-10 shows a depletion in the refractory elements relative to the volatile elements, which could be due to the formation of terrestrial planets in the Kepler-10 system. The average abundance pattern corresponds to $\sim$ 13 Earth masses, while the two known planets in Kepler-10 system have a combined $\sim$ 20 Earth masses. For two of the eight thick disc twins, however, no depletion patterns are found. Although our results demonstrate that several factors (e.g., planet signature, stellar age, stellar birth location and Galactic chemical evolution) could lead to or affect abundance trends with condensation temperature, we find that the trends give further support for the planetary signature hypothesis. 
\end{abstract}


\section{Introduction}

The technique of a strictly differential line-by-line analysis for measuring relative chemical abundances in stars with very high precision (0.01 dex, $\sim$2\%) has been further developed and applied to various cases over the past few years \citep{mel09,mel12,ram11,ram14,yon13,liu14,tm14,bia15,nis15,saf15}. This unprecedented precision has revealed subtle chemical differences in the photospheres of stars which have been interpreted as a signature of planet formation. \citet{mel09} demonstrated that the Sun exhibits a peculiar chemical pattern when compared to solar twins, namely, a depletion of refractory elements relative to volatile elements. They tentatively attributed this pattern to the formation of planets, especially rocky planets, in the solar system. In their scenario, refractory elements in the proto-solar nebula were locked up in the terrestrial planets. The remaining dust-cleansed gas was then accreted onto the Sun. In contrast, the typical solar twin did not form terrestrial planets efficiently, or dumped their proto-planetary nebulae, cleansed from refractories by planetary formation, so early that the stellar convection was still deep enough to dilute the disc gas to erase its chemical signature. Therefore, the Sun would exhibit a depletion in refractory elements relative to volatile elements when compared to most solar twins. \citet{cha10} confirm quantitatively that the depletion of refractories in the solar photosphere is possibly due to the depletion of a few Earth masses of rocky material.

This scenario, however, has been challenged by \citet{gh10} and \citet{adi14}. They argued that the observed trend between chemical abundances and condensation temperature (T$_{\rm c}$) could possibly be due to the differences in stellar ages rather than the presence of planets. \citet{nis15} conducted a high-precision differential abundance analysis for 21 solar twin stars in the solar neighbourhood with high signal-to-noise ratio (SNR $>$ 600) spectra. His results revealed abundance-age correlations for most elements. This indicates that chemical evolution in the Galactic disc might play an important role in the explanation of the trend between abundance and dust condensation temperature and must be considered when interpreting the results.

Another explanation for the peculiar solar composition is that the pre-solar nebula was radiatively cleansed from some of its dust by luminous hot stars in the solar neighbourhood before the formation of the Sun and its planets. This possibility is supported by the finding that the solar-age and rich open cluster M67 seems to have a chemical composition closer to the solar composition than most solar twins \citep{on11,on14}. A similar scenario was discussed by \citet{gai15}, who suggests that abundance-T$_{\rm c}$ correlations could be explained by dust-gas segregation in circumstellar discs.

The scenario put forward by \citet{mel09} makes a testable prediction that the host star of a system with terrestrial planets should also exhibit a depletion in refractory elements relative to volatile elements when compared to otherwise identical stars (i.e., stellar parameters, ages, birth locations). Therefore, in order to test this scenario, we need to conduct high precision chemical abundance studies of stars hosting terrestrial planets relative to similar other stars without such planets. Kepler-10 hosts two planets, Kepler-10b and Kepler-10c \citep{bat11}. \citet{dum14} reported that the mass of Kepler-10b is 3.33 $\pm$ 0.49 M$_\oplus$ with a density of 5.8 $\pm$ 0.8 g cm$^{-3}$, while the mass of Kepler-10c is 17.2 $\pm$ 1.9 M$_\oplus$ with a density of 7.1 $\pm$ 1.0 g cm$^{-3}$. \citet{dum14} characterized Kepler-10b and Kepler-10c as a hot Earth-like planet and a Neptune mass solid planet, respectively, although \citet{rog15} argued that Kepler-10c is likely to have a substantial volatile envelope and thus not rocky. The Kepler-10 system is thus a very suitable target to identify any chemical signatures of terrestrial planet formation. In particular, if the scenario presented by \citet{mel09} is correct, we should expect to find a deficiency of refractory elements relative to volatile elements in the photosphere of Kepler-10 when compared to other stars sharing similar stellar parameters but without known planets.

Here we present a strictly line-by-line differential abundance analysis of Kepler-10 and a sample of stellar twins to explore whether or not there is a chemical signature of terrestrial planet formation.

\section{Observations and data reduction}

We obtained high resolution and high SNR spectra with the Canada France Hawaii Telescope (CFHT), the Hobby-Eberly Telescope (HET) and the Magellan Clay Telescope.

We observed Kepler-10 with the Echelle SpectroPolarimetric Device for the Observation of Stars (ESPaDOnS) \citep{md03} at the Canada France Hawaii Telescope (CFHT) during June 2013. The spectral revolving power is 68,000 and the spectral range is 3800 -- 8900 \AA. In total eight spectra with exposures of 1700 s each were obtained. The individual frames were combined into a single spectrum with SNR $\approx$ 300 per pixel in most wavelength regions. A solar spectrum with even higher SNR ($\approx$ 500 per pixel) was obtained by observing the asteroid Vesta. The spectra were reduced with the CFHT data reduction tool 'Libre-Espirit' while the continuum normalizations were addressed with IRAF\footnote{IRAF is distributed by the National Optical Astronomy Observatory, which is operated by Association of Universities for Research in Astronomy, Inc., under cooperative agreement with National Science Foundation.}.

We also observed Kepler-10 with the High Resolution Spectrograph (HRS) \citep{tul98} on the Hobby-Eberly Telescope (HET) at McDonald Observatory during May 2011. A total integration time of 6.8 hours was needed to achieve SNR $>$ 350 per pixel. The spectrum has a spectral resolving power of 60,000 and covers 4100 to 7800 \AA, with a gap of about 100 \AA\, around 6000 \AA. A solar spectrum with higher spectral resolution (R = 120,000) and higher SNR ($\approx$ 500 per pixel) was obtained by observing the asteroid Iris. The HRS-HET data were reduced using IRAF's echelle package.

We selected fourteen stars identified as Kepler-10 stellar twins, based on the similarity of their stellar parameters (T$_{\rm eff}$, $\log g$, [Fe/H]) to those of Kepler-10, using an updated version of the stellar parameter catalog of \citet{rm05} and from the sample by \citet{ben14}. The comparison star sample was chosen randomly such that any individual star was not necessarily included in planet search programs. Those "Kepler-10 twins" were observed using the Magellan Inamori Kyocera Echelle (MIKE) spectrograph \citep{ber03} during two runs: June 2014 and June 2015. The spectrograph delivers wavelength coverage from about 3300 to 5000 \AA\, (blue arm) and 4900 to 9400 \AA\, (red arm) at a spectral resolving power of 83,000 and 65,000, respectively, the SNR exceeded 300 per pixel at 6000 \AA. A solar spectrum using the asteroid Vesta was obtained each night in the first run. We reduced the spectra with standard procedures which include bias subtraction, flat-fielding, scattered-light subtraction, 1D spectral extraction, wavelength calibration, and continuum normalization, with IRAF.

Our thick disc twins were not observed with the northern telescopes, nor was it possible to observe Kepler-10 from the southern Magellan site. This limits our strictly differential study to the use of the solar-spectrum observations as a test calibration. We also carried out a number of tests which ensure that our results are not compromised by the use of different spectroscope/telescope combinations. The most important of these tests are described later in this paper.

\section{Stellar atmospheric parameters}

The line list employed in our analysis was adopted mainly from \citet{asp09} and complemented with additional unblended lines from \citet{ben05} and \citet{nev09}; in a differential abundance analysis the accuracy of the \textit{gf} values does not influence the results. Equivalent widths (EWs) were measured using the ARES code \citep{sou07} for most lines. The equivalent widths for C, O, Mg, Al, S, Mn, Cu and Zn (i.e., elements with fewer lines) were measured manually with the \textit{splot} task in IRAF. Weak ($<$ 5 m\AA) and strong ($>$ 110 m\AA) lines were excluded from the analysis. The atomic line data adopted for the abundance analysis are listed in Table A1. We emphasize that in a differential analysis such as ours, the atomic data have essentially no influence on the results since Kepler-10 and its twins have very similar stellar parameters.

We performed a 1D, local thermodynamic equilibrium (LTE) abundance analysis using the 2013 version of MOOG \citep{sne73,sob11} with the ODFNEW grid of Kurucz model atmospheres \citep{cas03}. Stellar parameters were obtained by forcing excitation and ionization balance of Fe\,{\sc i} and Fe\,{\sc ii} lines on a line-by-line basis relative to the Sun. The adopted parameters for the Sun were $T_{\rm eff} = 5777$\,K, $\log g = 4.44$, [Fe/H] = 0.00, $\xi_{\rm t}$ = 1.00 km\,s$^{-1}$. The stellar parameters of Kepler-10 and its stellar twins were established separately using an automatic grid searching method described by \citet{liu14}. The best combination of $T_{\rm eff}$, $\log g$, [Fe/H] and $\xi_{\rm t}$, minimizing the slopes in [Fe\,{\sc i}/H] versus excitation potential and reduced equivalent width as well as the difference between [Fe\,{\sc i}/H] and [Fe\,{\sc ii}/H], is obtained from a successively refined grid of stellar atmospheric models. The final solution was obtained when the grid step-size decreased to $\Delta T_{\rm eff} = 1$ K, $\Delta \log g = 0.01$ and $\Delta \xi_{\rm t}$ = 0.01 km\,s$^{-1}$. We also required the derived average [Fe/H] to be consistent with the adopted model atmospheric value. Lines whose abundances departed from the average by $> 2.5\sigma$ were clipped.

\begin{figure}
\centering
\includegraphics[width=\columnwidth]{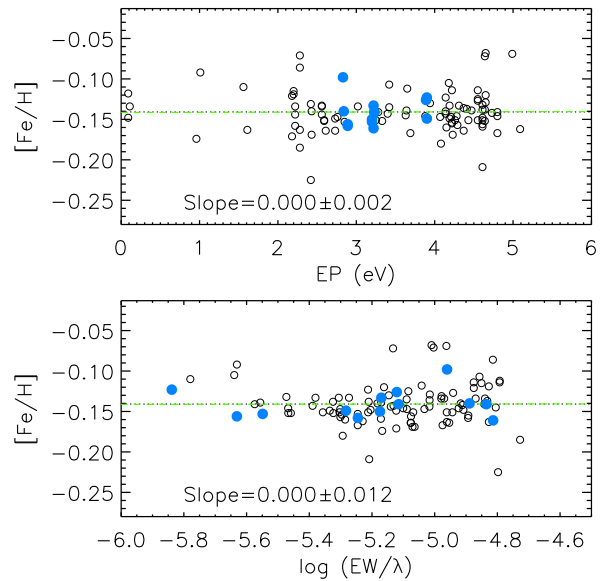}
\caption{Top panel: [Fe/H] of Kepler-10 derived on a line-by-line basis with respect to the Sun as a function of lower excitation potential; open circles and blue filled circles represent Fe \,{\sc i} and Fe \,{\sc ii} lines, respectively. The black dotted line shows the location of mean [Fe/H], the green dashed line represents the best fit to the data. Bottom panel: same as in the top panel but as a function of reduced equivalent width.}
\label{fig0}
\end{figure}

Figure \ref{fig0} shows an example of determining the stellar parameters of Kepler-10. The adopted stellar parameters satisfy the excitation and ionization balance in a differential sense. The best fit $\pm$ 1$\sigma$ for the [Fe/H] versus excitation potential (EP) roughly corresponds to an error in T$_{\rm eff}$ of 10 K, similarly for the reduced EW ($\log$ (EW/$\lambda$), which corresponds to an error of $\sim$ 0.02 - 0.03 km\,s$^{-1}$ in $\xi_{\rm t}$. The abundance difference in Fe \,{\sc i} and Fe \,{\sc ii} = 0.000 $\pm$ 0.006, which constrains $\log g$ to a precision of 0.02 - 0.03.

For the fourteen stellar twins, differential stellar parameters were also obtained by the line-by-line differential analysis as described before, but relative to Kepler-10 rather than the Sun, i.e., Stellar twins (Magellan) $-$ Kepler-10 (HET or CFHT). The adopted initial parameters for Kepler-10 were $T_{\rm eff} = 5700$\,K, $\log g = 4.35$, [Fe/H] = $-$0.15, $\xi_{\rm t}$ = 1.00 km\,s$^{-1}$, taken from the Kepler-10 analysis relative to the Sun. We emphasize that the absolute values are not crucial for our differential abundance analysis. We did not consider $\alpha$ enhancements in the thick disc stars in the model atmospheres but this does not affect our results, in particular not in the differential study of Kepler-10 relative to its thick disc twins. We assume that the stellar spectrum is defined solely by the stellar parameters $T_{\rm eff}$, $\log g$, $\xi_{\rm t}$ and abundances, i.e., other individual stellar parameters, e.g. describing stellar activity is not considered in this study.

The final adopted atmospheric parameters of Kepler-10 and its stellar twins are listed in Table \ref{t:para}. The uncertainties in the stellar parameters were derived with the method described by \citet{eps10} and \citet{ben14}, which accounts for the co-variances between changes in the stellar parameters and the differential abundances. Excellent precision was achieved due to the strictly differential method, which should greatly reduce the systematic errors from atomic line data and shortcomings in the 1D LTE modelling of the stellar atmospheres and spectral line formation (e.g., \citealp{asp05}).

\begin{table*}
\caption{Stellar parameters of Kepler-10 and its stellar twins.}
\label{t:para}
\begin{tabular*}{\textwidth}{lcccccccc}
\hline
Object & $T_{\rm eff}$ & $\log g$ & $\xi_{\rm t}$ & [Fe/H] & Probability$^d$ & Probability$^d$ & Population & Age \\
 & (K) & & (km\,s$^{-1}$) &  & Thin disc (\%) & Thick disc (\%) & & (Gyr) \\
\hline
Kepler-10$^a$ & 5697$\pm$10 & 4.40$\pm$0.03 & 0.98$\pm$0.02 & $-$0.141$\pm$0.009 & 3 & 96 & thick & 8.4$\pm$1.0\\
Kepler-10$^b$ & 5695$\pm$13 & 4.38$\pm$0.04 & 0.96$\pm$0.03 & $-$0.143$\pm$0.015 & 3 & 96 & thick & 9.0$\pm$1.1\\
\hline
HD 88084$^c$  & 5780$\pm$13 & 4.40$\pm$0.03 & 1.05$\pm$0.03 & $-$0.091$\pm$0.012 & 98 & 2 & thin & 5.8$\pm$1.4\\
HD 115382$^c$  & 5776$\pm$12 & 4.38$\pm$0.03 & 1.05$\pm$0.03 & $-$0.089$\pm$0.010 & 92 & 8 & thin & 6.7$\pm$1.4\\
HD 126525$^c$  & 5687$\pm$19 & 4.50$\pm$0.03 & 1.00$\pm$0.04 & $-$0.063$\pm$0.014 & 91 & 9 & thin & 2.6$\pm$1.5\\
HD 117939$^c$  & 5729$\pm$13 & 4.45$\pm$0.03 & 1.00$\pm$0.03 & $-$0.176$\pm$0.016 & 60 & 40 & thin & 4.7$\pm$1.5\\
HIP 113113$^c$ & 5706$\pm$13 & 4.42$\pm$0.04 & 0.96$\pm$0.03 & $-$0.071$\pm$0.012 & 58 & 42 & thin & 5.6$\pm$1.6\\
\hline
HD 117126$^c$  & 5779$\pm$18 & 4.25$\pm$0.04 & 1.10$\pm$0.03 & $-$0.032$\pm$0.015 & 55 & 44 & thick & 8.2$\pm$0.5\\
HD 115231$^c$  & 5708$\pm$15 & 4.43$\pm$0.04 & 1.00$\pm$0.04 & $-$0.095$\pm$0.015 & 38 & 61 & thick & 5.8$\pm$1.6\\
HD 106210$^c$  & 5701$\pm$13 & 4.40$\pm$0.04 & 0.96$\pm$0.03 & $-$0.131$\pm$0.014 & 18 & 80 & thick & 7.0$\pm$1.6\\
HIP 109821$^c$ & 5772$\pm$14 & 4.32$\pm$0.03 & 1.06$\pm$0.03 & $-$0.087$\pm$0.011 & 17 & 82 & thick & 7.7$\pm$0.8\\
HD 87320$^c$   & 5666$\pm$12 & 4.40$\pm$0.03 & 0.93$\pm$0.03 & $-$0.149$\pm$0.010 & 11 & 87 & thick & 7.5$\pm$1.4\\
HIP 96124$^c$  & 5636$\pm$13 & 4.41$\pm$0.04 & 0.93$\pm$0.03 & $-$0.197$\pm$0.012 & 9 & 90 & thick & 8.2$\pm$1.6\\
HIP 101857$^c$ & 5798$\pm$14 & 4.34$\pm$0.03 & 1.06$\pm$0.03 & $+$0.029$\pm$0.012 & 2 & 96 & thin$^e$ & 6.4$\pm$0.8\\
HIP 9381$^c$  & 5734$\pm$14 & 4.39$\pm$0.04 & 1.02$\pm$0.03 & $-$0.238$\pm$0.012 & 1 & 96 & thick & 8.0$\pm$1.7\\
HIP 99224$^c$  & 5754$\pm$18 & 4.27$\pm$0.04 & 1.03$\pm$0.04 & $-$0.004$\pm$0.014 & 0 & 96 & thick & 8.2$\pm$0.6\\
\hline
\end{tabular*}
\begin{minipage}{\textwidth}
$^a$ Parameters derived with HET data. $^b$ Parameters derived with CFHT data.\\
$^c$ Parameters derived using Kepler-10 (HET) as the reference.\\
$^d$ Probabilities calculated based on kinematics \citep{ram13}.\\
$^e$ HIP 101857 is assigned to the thin disc because of its abundance pattern rather than kinematics.
\end{minipage}
\end{table*}

\section{Results}

\subsection{Elemental abundances}

Having established the stellar parameters for Kepler-10 and its stellar twins, we derived chemical abundances relative to the Sun for an additional 17 elements from atomic lines: C, O, Na, Mg, Al, Si, S, Ca, Sc, Ti, V, Cr, Mn, Co, Ni, Cu and Zn. We also obtained differential abundances of Kepler-10's stellar twins relative to Kepler-10. Hyperfine-structure splitting (HFS) was considered for Sc, V, Cr, Mn and Cu using data from \citet{kur95}. Departures from LTE were considered for the 777 nm oxygen triplet lines according to \citet{ram07} and the typical size of the correction is $\approx$ $-$0.01 dex. The errors in the differential abundances were calculated following the method of \citet{eps10}: the standard errors in the mean abundances, as derived from the different spectral lines, were added in quadrature to the errors introduced by the uncertainties in the atmospheric parameters. Most derived elemental abundances have uncertainties $\leq$ 0.02 dex, which further underscores the advantages of a strictly differential analysis. Indeed, when considering all elements, the average uncertainty is only 0.014 $\pm$ 0.002 ($\sigma$ = 0.006) for Kepler-10 relative to the Sun.

\begin{figure}
\centering
\includegraphics[width=\columnwidth]{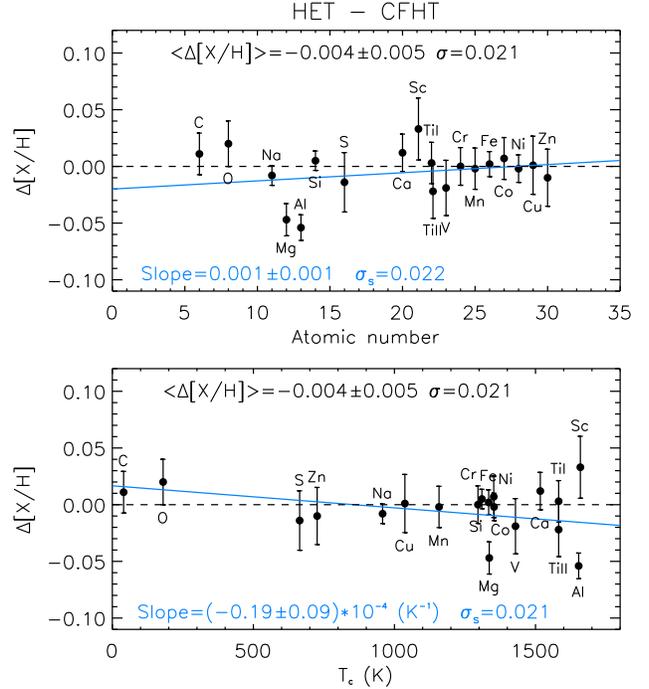}
\caption{Top panel: Abundance differences for Kepler-10 from two different telescopes, $\Delta$[X/H] (HET $-$ CFHT), versus atomic number; the blue solid line represents the linear fit to the data; $\sigma_s$ is the dispersion about the linear fit. Bottom panel: Abundance differences as a function of condensation temperature T$_{\rm c}$.}
\label{fig1}
\end{figure}

We first compare the abundances of Kepler-10 as derived from HET and CFHT spectra (Figure \ref{fig1}). The values of T$_{\rm c}$ (specifically 50\% condensation temperature for a solar-composition mixture) are given by \citet{lod03}. The average abundance difference $\Delta$[X/H] (HET $-$ CFHT) is $-$0.004 $\pm$ 0.005 ($\sigma$ = 0.021), consistent with zero. We perform a least-square linear fit weighted by the errors in abundances while the uncertainties of the fitting are calculated considering the chi-square merit function and the relative derivatives\footnote{We applied the same manner to all the following linear fits.}. We note that there is a slight negative trend between $\Delta$[X/H] versus T$_{\rm c}$ with a slope of ($-$0.19 $\pm$ 0.09) $\times$ 10$^{-4}$ K$^{-1}$, which is mainly driven by the two volatile elements, C and O. We adopt the results derived from the HET spectra. This choice does not affect our conclusions since the differences between HET and CFHT data are very small. We do explore the effects of using CFHT data below.

\begin{figure}
\centering
\includegraphics[width=\columnwidth]{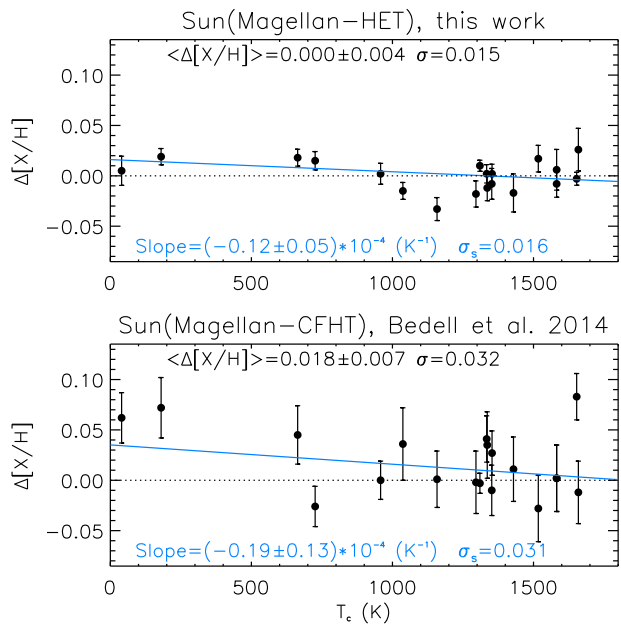}
\caption{Abundance differences $\Delta$[X/H] versus condensation temperature T$_{\rm c}$ for solar spectra obtained with different instruments (Magellan $-$ HET) for this work (top panel) and for \citet{bed14} (Magellan $-$ CFHT) (bottom panel). The blue solid lines represent the linear fit to our data (top panel) and the \citet{bed14} data (bottom panel), $\sigma_s$ is the dispersion about the linear fit.}
\label{fig2}
\end{figure}

Another issue regarding systematic offsets that we need to consider carefully is whether the choice of the spectrograph affects the results. According to \citet{bed14}, systematic offsets may be introduced when comparing the results using spectra obtained from different instruments or when measurements are not performed consistently. While the Kepler-10 differential abundances were measured based on HET spectra, differential abundances for the Kepler-10 stellar twins were measured based on Magellan spectra. Therefore, it is crucial to check whether any systematic offsets exist. In Figure \ref{fig2}, we plot $\Delta$[X/H] as a function of T$_{\rm c}$ derived from solar spectra obtained with different instruments (Sun (Magellan $-$ HET)). In that figure, we also include a comparison of the Sun (Magellan $-$ CFHT) from \citet{bed14}. The average difference in our $\Delta$[X/H] is 0.000 $\pm$ 0.004 ($\sigma$ = 0.015) and the slope of the linear fit is ($-$0.12 $\pm$ 0.05) $\times$ 10$^{-4}$ K$^{-1}$. The systematic offsets in our work are much smaller than that in \citet{bed14}. One possible reason for this difference is that in \citet{bed14}, the normalization of spectra and the measurement of EWs involved not only different instruments, but also different investigators. In this work, the entire analysis was done consistently by one person using the same approach, minimizing the possible systematic errors introduced by comparing the results based on different instruments.

\begin{figure*}
\centering
\includegraphics[width=0.98\hsize]{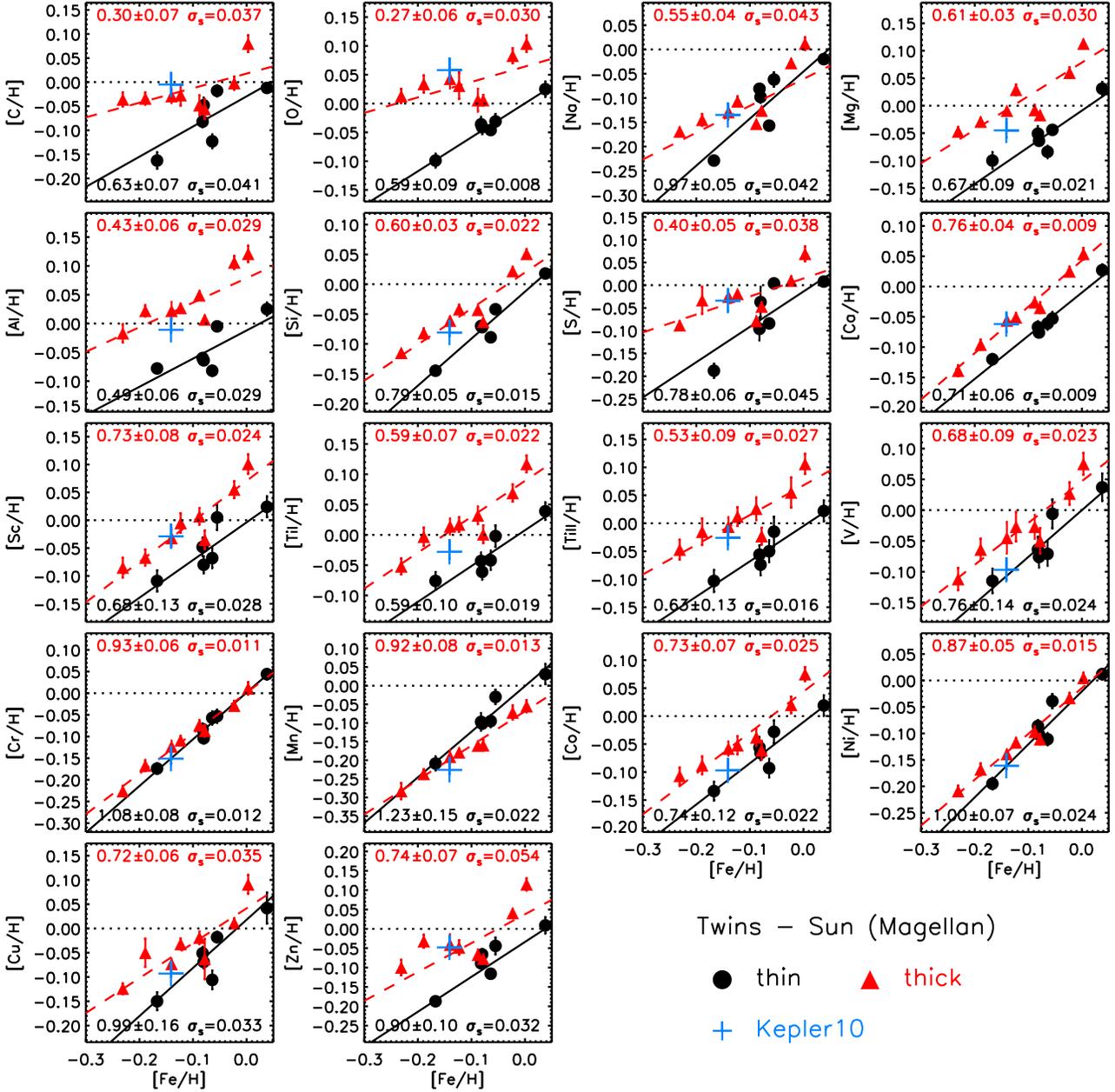}
\caption{[X/H] versus [Fe/H] for various elements for the "Kepler-10 twins" (Twins $-$ Sun (Magellan)). Linear fits for the thin disc counterparts (black circles) and thick disc (red triangles) twins are overplotted, $\sigma_s$ is the dispersion about the linear fit. The location of Kepler-10 is marked (Kepler-10 $-$ Sun (HET), blue crosses). The size of the crosses are corresponding to the error bars in [X/H] and [Fe/H].}
\label{fig3}
\end{figure*}

The discovery paper by \citet{bat11} reported [Fe/H] = $-$0.15 $\pm$ 0.04 for Kepler-10. We confirm that Kepler-10, with [Fe/H] = $-$0.141 $\pm$ 0.009, is metal-poor relative to the Sun\footnote{Recently, a very similar metallicity of [Fe/H] = $-$0.14 $\pm$ 0.02 was presented by \citet{san15}}. Kepler-10 is also older than the Sun with an age of 8.4 $\pm$ 1.0 Gyr from our derivation, see below. The total space velocity ($V_{\rm tot} = U^2+V^2+W^2)^{1/2}$) of Kepler-10 is 97.0 km\,s$^{-1}$ and the kinematic probability of being from the thick disc is 96\% \citep{dum14}. Therefore direct comparisons of Kepler-10 to the Sun is not adequate. Kepler-10 should be compared against stars of similar metallicity and belonging to the same stellar population. For the fourteen Kepler-10 stellar twins without known planets, we show the distribution in [Fe/H] and [X/H] in Figure \ref{fig3}. We calculated the Galactic space velocities U, V, W of our sample stars using data from \textit{Simbad} database with the equations given by e.g., \citet{js87}. We derived the associated probabilities of thin/thick disc membership based on the algorithm described by \citet{ram07,ram13}. We computed the stellar ages using the stellar parameters and their errors as given in Table \ref{t:para}, placing them on a T$_{\rm eff}$ - $\log g$ plane, and comparing these locations with the theoretical isochrones of the Yonsei-Yale group (e.g., \citealp{yi01,kim02}). Details of our age determination technique are provided in \citet{ram14}.

We have three criteria for thick disc membership: kinematic probability $>$ 60\%, age $>$ 7 Gyr and chemical similarity with thick disc stars. All the eight thick disc twins fulfil at least two of these criteria (see Table \ref{t:para}). The remaining program stars are likely thin disc members. Regarding the latter criterion, it is evident from Figure \ref{fig3} (and previous work by \citealp{red06,ben14}) that thin and thick disc stars lie on different and well-defined trends, although there are also some objects that exhibit thick disc kinematics but thin disc abundances \citep{red06}. In the present work, we are searching for subtle chemical abundance differences among thick disc stars, so it is important that these comparison stars have thick disc chemical abundances. 

\subsection{$\Delta$[X/H] - T$_{\rm c}$ correlations}

The [X/H] ratios confirm that Kepler-10 is a thick disc object and its relatively old age further supports this. 
Therefore, we will compare Kepler-10 against its thick disc stellar twins in order to compensate for effects of Galactic chemical evolution. As is seen directly from Figure \ref{fig3}, the abundances of Kepler-10 show a systematical pattern relative to the linear fits in the panel that presumably display the Galactic chemical evolution as relations between [X/H] and [Fe/H]. E.g., for the five elements with the lowest condensation temperatures, C, O, S, Zn and Na, the blue crosses representing Kepler-10 in the panels of the figure are situated on or above the redline, while for the eight elements with the highest condensation temperature, Mg, Co, Ni, V, Ca, Ti, Al and Sc, Kepler-10 is located on or below the redline. It is easy to demonstrate that if we assume that the real abundances of Kepler-10 would be on the line and the observed locations are reflecting independent errors symmetrically distributed (i.e. with equally probable departures in positive or negative directions), the chance of obtaining this systematic effect with T$_{\rm c}$ by mere chance is less than 1\%. In view of the fact that the present study was initiated when the super-Earths of Kepler-10 had been discovered in order to test the planetary signature of the abundance - T$_{\rm c}$ relation, the systematics of Figure \ref{fig3} is in itself a striking confirmation, indicating that this interpretation must be favoured relative to e.g. Chemical evolution effects.

\begin{figure*}
\centering
\includegraphics[width=0.98\hsize]{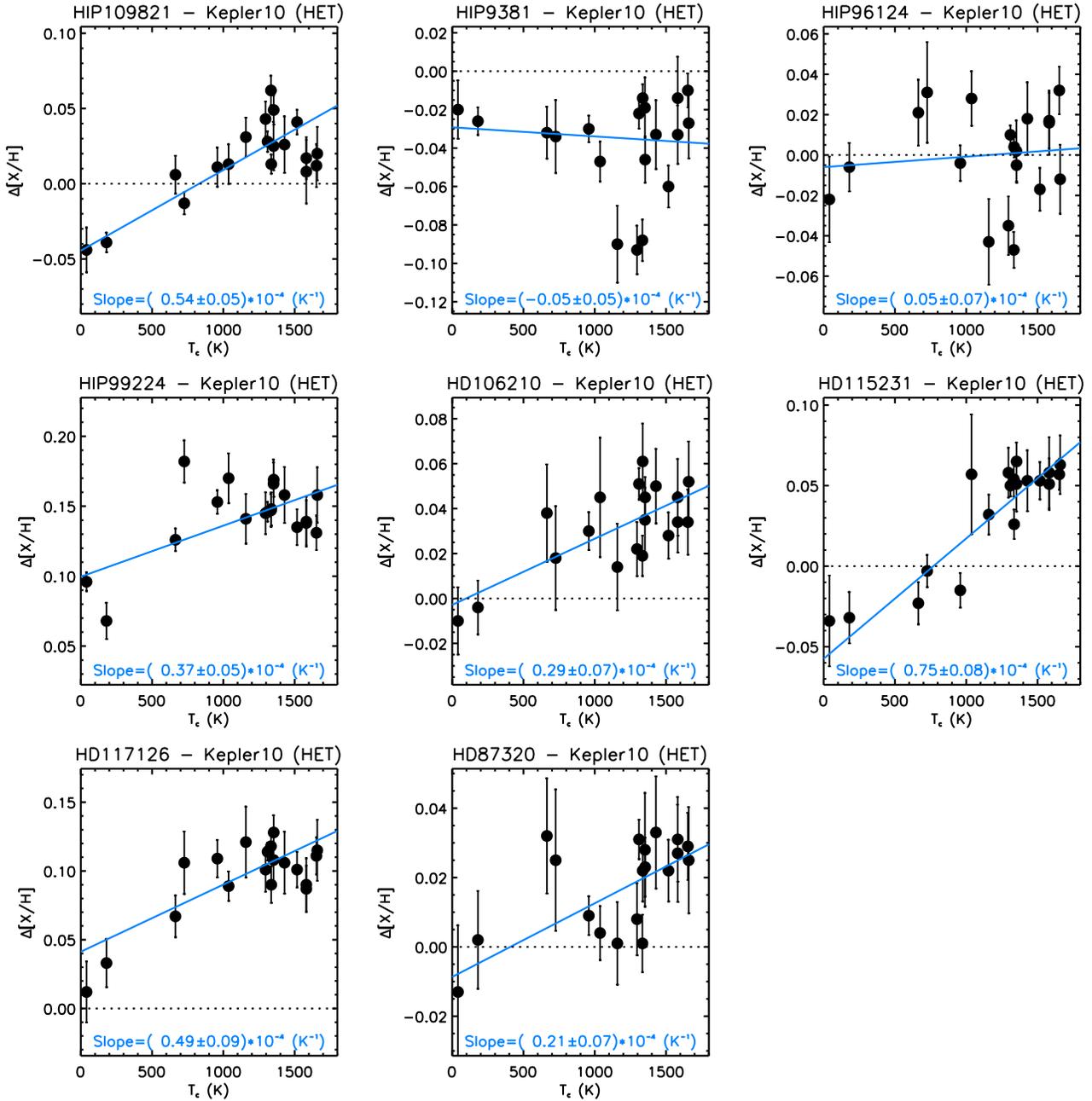}
\caption{Abundance differences $\Delta$[X/H] versus condensation temperature T$_{\rm c}$ for all the thick disc stellar twins relative to Kepler-10. The blue solid lines show the single linear fit to the results.}
\label{fig4}
\end{figure*}

To improve the precision further, we derive strictly differential abundances $\Delta$[X/H] for the eight likely thick disc stellar twins relative to Kepler-10 in Figure \ref{fig4} rather than relative to the Sun as the case for Figure \ref{fig3}. 
We find that a single linear fit provides an appropriate representation of the $\Delta$[X/H] - T$_{\rm c}$ correlation when comparing the thick disc twins to Kepler-10. Our results demonstrate that the $\Delta$[X/H] - T$_{\rm c}$ trends could vary from star to star, as reported by \citet{nis15}. Five stars (HIP 109821, HIP 99224, HD 106210, HD 115231 and HD 117126) show positive slopes for the single linear fitting but the trends are driven mainly by the abundances of the most volatile elements C and O. HD 87320 shows a positive slope as well but with much larger scatter around the best fit. HIP 9381 and HIP 96124 show large scatters around the zero-slopes with three elements as outliers (Cr, Mn and Fe). These outlier elements could be due to the impact of GCE since these two stars are the most metal-poor and those three elements (Cr, Mn and Fe) exhibit the steepest slopes for the [X/H] vs. [Fe/H] in Figure \ref{fig3}. 

\begin{figure}
\centering
\includegraphics[width=\columnwidth]{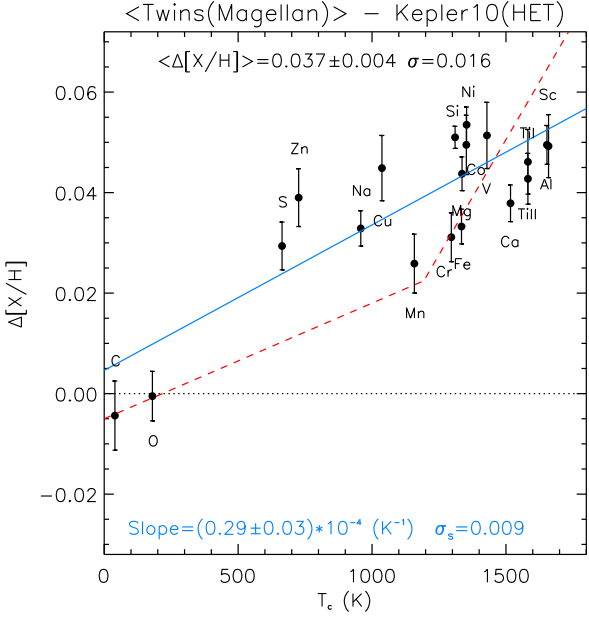}
\caption{Average abundance differences $\Delta$[X/H] versus condensation temperature T$_{\rm c}$ for the eight thick disc stellar twins relative to Kepler-10. The blue solid line represents the linear fit to the data, $\sigma_s$ is the dispersion about the linear fit and the red dashed line is the fit from \citet{mel09} for solar twins $-$ Sun, normalized to $\Delta$[C/H].}
\label{fig5}
\end{figure}

We average the results of the differential abundances $\Delta$[X/H] for these eight thick disc stellar twins (i.e. $<$Thick disc twins (Magellan)$>$ $-$ Kepler-10 (HET)) and show the result in Figure \ref{fig5}. As seen already directly from Figure \ref{fig3}, Kepler-10 shows a depletion of refractory elements relative to the volatile elements when compared to the average of all the thick disc stellar twins. The average difference is $\Delta$[X/H] = 0.037 $\pm$ 0.004 ($\sigma$ = 0.016). A linear fit to the data has a gradient of (0.29 $\pm$ 0.03) $\times$ 10$^{-4}$ K$^{-1}$, corresponding to a $>$ 9$\sigma$ significance. Although the trend is mainly driven by C and O, a significant trend (slope = (0.17 $\pm$ 0.04) $\times$ 10$^{-4}$ K$^{-1}$) is also present when excluding these two elements. Table \ref{t:abun} lists the adopted elemental abundances and associated uncertainties of Kepler-10 and the average of its thick disc stellar twins. In addition, Table A2 lists all the derived elemental abundances and associated uncertainties of each program star with relative to Kepler-10.

\begin{table}
\setlength{\tabcolsep}{3.5pt}
\caption{[X/H] for Kepler-10 and the average of its thick disc stellar twins.}
\label{t:abun}
\begin{tabular}{crrc}
\hline
Element & Kepler-10$^a$ & Kepler-10$^b$ & $<$Thick disc twins$>^c$ \\
\hline
C  & $-$0.005$\pm$0.015 & $-$0.016$\pm$0.011 & $-$0.004$\pm$0.006\\
O  &  0.058$\pm$0.010 & 0.038$\pm$0.017 & $-$0.001$\pm$0.005\\
Na & $-$0.135$\pm$0.007 & $-$0.127$\pm$0.005 & 0.033$\pm$0.003\\
Mg & $-$0.045$\pm$0.013 & 0.002$\pm$0.006 & 0.044$\pm$0.004\\
Al & $-$0.011$\pm$0.005 & 0.043$\pm$0.010 & 0.050$\pm$0.004\\
Si & $-$0.081$\pm$0.006 & $-$0.086$\pm$0.006 & 0.051$\pm$0.002\\
S  & $-$0.034$\pm$0.022 & $-$0.020$\pm$0.014 & 0.029$\pm$0.005\\
Ca & $-$0.062$\pm$0.013 & $-$0.074$\pm$0.010 & 0.038$\pm$0.004\\
Sc & $-$0.029$\pm$0.018 & $-$0.062$\pm$0.020 & 0.049$\pm$0.006\\
Ti\,{\sc i}  & $-$0.028$\pm$0.012 & $-$0.031$\pm$0.014 & 0.043$\pm$0.005\\
Ti\,{\sc ii} & $-$0.026$\pm$0.012 & $-$0.004$\pm$0.021 & 0.046$\pm$0.006\\
V  & $-$0.097$\pm$0.016 & $-$0.078$\pm$0.018 & 0.051$\pm$0.007\\
Cr & $-$0.151$\pm$0.012 & $-$0.151$\pm$0.011 & 0.031$\pm$0.005\\
Mn & $-$0.226$\pm$0.011 & $-$0.224$\pm$0.015 & 0.026$\pm$0.006\\
Fe & $-$0.141$\pm$0.009 & $-$0.143$\pm$0.015 & 0.033$\pm$0.003\\
Co & $-$0.097$\pm$0.013 & $-$0.104$\pm$0.013 & 0.050$\pm$0.006\\
Ni & $-$0.161$\pm$0.007 & $-$0.159$\pm$0.010 & 0.054$\pm$0.004\\
Cu & $-$0.093$\pm$0.006 & $-$0.094$\pm$0.025 & 0.045$\pm$0.007\\
Zn & $-$0.048$\pm$0.006 & $-$0.038$\pm$0.025 & 0.039$\pm$0.006\\
\hline
\end{tabular}
$^a$ [X/H] derived with HET data, relative to the Sun.\\
$^b$ [X/H] derived with CFHT data, relative to the Sun.\\
$^c$ $\Delta$[X/H] derived with respect to Kepler-10 (HET).
\end{table}

\begin{figure}
\centering
\includegraphics[width=0.98\hsize]{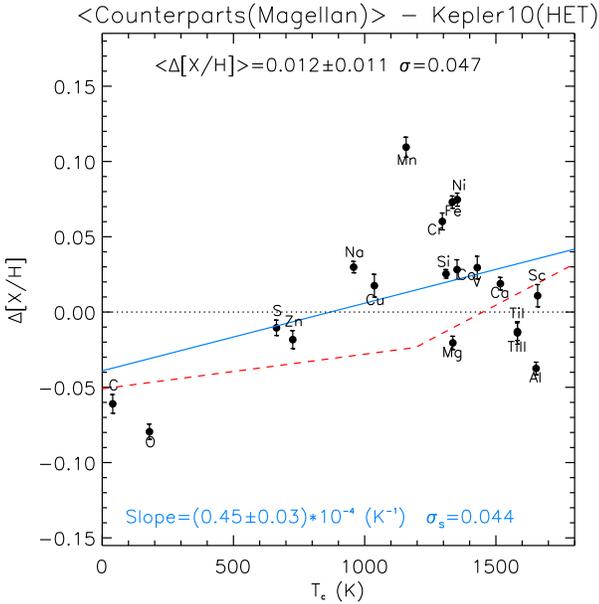}
\caption{Average abundance differences $\Delta$[X/H] versus condensation temperature T$_{\rm c}$ for the thin disc counterparts relative to Kepler-10. The blue solid line represents the linear fit to the data, $\sigma_s$ is the dispersion about the linear fit and the red dashed line is the fit from \citet{mel09} for solar twins $-$ Sun.}
\label{fig6}
\end{figure}

Although the classification of Kepler-10 as a thick disc star, on the basis of its kinematics, age and abundance pattern, one may ask how the abundance pattern relative to T$_{\rm c}$ would look if compared with its thin disc counterpart stars, instead. In Figure \ref{fig6} we display the differences between the mean of [X/H] for the thin disc stars and Kepler-10, again plotted vs. T$_{\rm c}$. A linear fit to the data has a gradient of (0.45 $\pm$ 0.03) $\times$ 10$^{-4}$ K$^{-1}$, while the dispersion about the linear fit ($\sigma_s$) is 0.044 dex. The trend excluding C and O has a gradient to be ($-$0.12 $\pm$ 0.04) $\times$ 10$^{-4}$ K$^{-1}$ with 2.6$\sigma$ significance. We see that the systematic slope of the relation still prevails, but that it is now very much dependent on C and O; the relation for the rest of the elements show a characteristic peak, corresponding to elements with T$_{\rm c}$ $\sim$ 1200 K. This demonstrates that Galactic chemical evolution partly masks the effects of dust-depletion on the abundance pattern.

As a further check of our results, we also repeated the analysis using the CFHT Kepler-10 spectrum and analyzed the stellar twins with respect to that spectrum. The results are very similar to those presented in Figures. 5-7.

\section{Discussion}

\begin{figure}
\centering
\includegraphics[width=\columnwidth]{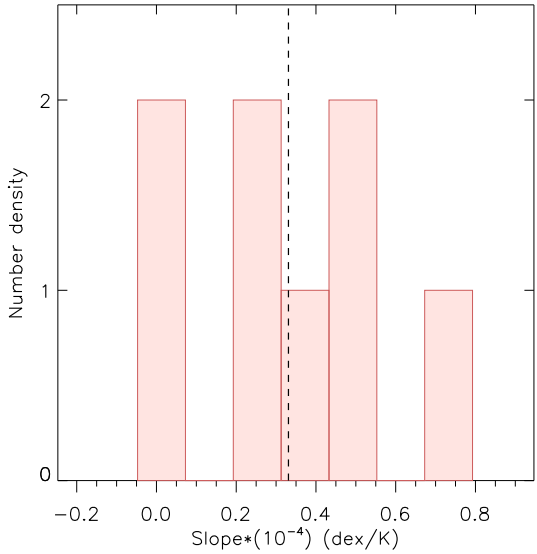}
\caption{Histogram of the slopes when applying a single linear fit to $\Delta$[X/H] - T$_{\rm c}$ correlations for the eight thick disc stellar twins. The black dashed vertical line represents the location of the mean value of $\Delta$[X/H] vs. T$_{\rm c}$ slopes.}
\label{fig7}
\end{figure}

As shown in Figure \ref{fig5}, there is a deficiency of refractory elements relative to volatile elements in the photosphere of the terrestrial planet host Kepler-10 when compared to the average results of its thick disc stellar twins without known planets. Using the current size of the convective zone of Kepler-10 (0.08 M$_{\odot}$, \citealp{sie00}), the abundance pattern corresponds to at least 13 Earth masses of rocky material \citep{cha10} which is comparable to the total mass of planets (20 Earth masses) in the Kepler-10 system. 
Therefore the differences in chemical composition between Kepler-10 and its thick disc stellar twins could be attributed to the formation of terrestrial planets in the Kepler-10 system, but this requires that the life time of the proto-planetary disc was long enough to not deliver its dust-cleansed gas until the convection zone of the star reached its present depth. As we mentioned before, even for the thick disc twins which share similar stellar parameters and ages with Kepler-10, the $\Delta$[X/H] - T$_{\rm c}$ correlations still vary star to star. In order to investigate this further, we show the histogram of the slopes for the single linear fitting of the T$_{\rm c}$ trends for the eight thick disc stars in Figure \ref{fig7}. The slopes exhibit a broad distribution. We note that two (HIP 9381 and HIP 96124) of the thick disc stars do not show any apparent trends, which complicates the scenario of the chemical signatures of terrestrial planets. If the $\Delta$[X/H] - T$_{\rm c}$ trends do reflect planet formation, those two stars could be conjectured to also harbour terrestrial planets that have not yet been detected. The first one (HIP 9381) has been observed multiple times with HARPS yet no results have been published. It is also probable that other factors play a role in determining the detailed chemical composition of those stars. 

\begin{figure}
\centering
\includegraphics[width=\columnwidth]{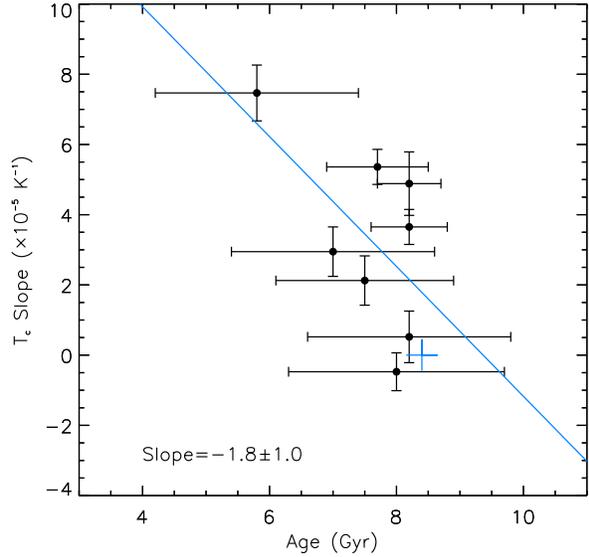}
\caption{Gradient for a single linear fit to $\Delta$[X/H] vs. T$_{\rm c}$ slopes as a function of stellar ages for thick disc stars. $\Delta$[X/H] were measured using Kepler-10 as a reference. The blue solid line represents the linear fit for the thick disc stars. The location of Kepler-10 is marked with blue cross.}
\label{fig8}
\end{figure}

\citet{adi14} and \cite{nis15} proposed that the trends between chemical abundance and condensation temperature (T$_{\rm c}$) could be due to the differences in the stellar ages. We plot the differential abundances $\Delta$[X/H] - T$_{\rm c}$ slopes versus stellar ages in Figure \ref{fig8}. A linear fit to the data is over-plotted (each data point is given equal weight). 
The gradient is $-$1.8 $\pm$ 1.0 for thick disc twins, using Kepler-10 as a reference. The negative slope is likely driven by the one star younger than 6 Gyr. Without that object, the diagram is a scatter plot such that age alone can not explain the chemical behaviour. We note that for most thick disc twins, although they have similar ages, the $\Delta$[X/H] - T$_{\rm c}$ slopes can vary by $\sim$ 6 $\times$ 10$^{-5}$ K$^{-1}$. Therefore, we emphasize again that age alone can not explain the chemical patterns found in Figure \ref{fig4} and Figure \ref{fig5}.

\begin{figure}
\centering
\includegraphics[width=\columnwidth]{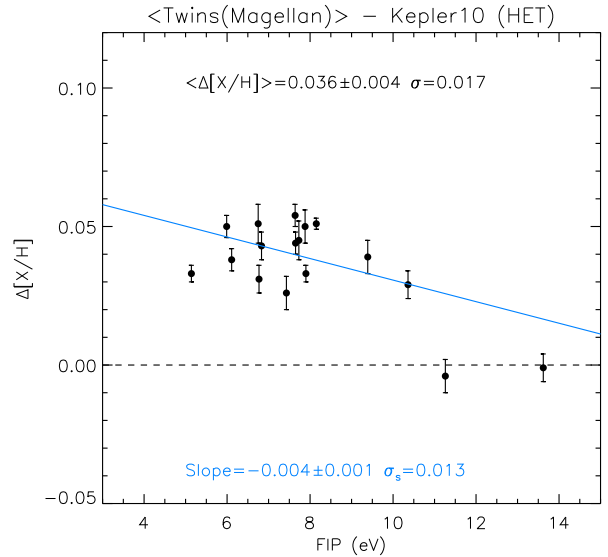}
\caption{Differential chemical abundances $\Delta$[X/H] as a function of first ionization potential (FIP). The blue solid line represents the linear fit to the data, $\sigma_s$ is the dispersion about the linear fit.}
\label{fig9}
\end{figure}

It has since long been known that the solar upper atmosphere and wind abundances are affected by anomalies, with respect to the photosphere, in that the elements with a high first ionization potential (such as Ne and Ar) are depleted relative to those with low potentials (e.g., Fe, Mg, Si, see \citealp{fl00}). Effects of this kind could possibly occur differentially in stellar photospheres, and might mimic the abundance correlations with condensation temperature. We have explored this by examining the relation between $\Delta$[X/H] and the first ionization potential (FIP) in Figure \ref{fig9}. Although a correlation is apparent, its significance (4$\sigma$) is much less than for the T$_{\rm c}$ trend. Figure \ref{fig9} might be just as well considered as providing two clumps: C and O, the rest of the elements, respectively. A similar phenomenon was also reported by \citet{ram10} for the \citet{mel09} and \citet{ram09} solar twin data sets. We performed a Spearman correlation test of the abundance differences vs. T$_{\rm c}$ and FIP. The Spearman correlation coefficient is $r_S$ = $+$0.68 when using T$_{\rm c}$, but only $-$0.32 for the FIP. The probability of a correlation arising by chance is 0.2\% for T$_{\rm c}$, while the probability of the correlation with FIP arise by chance is 20.6\%. We emphasize that there is no convincing physical scenario to explain the FIP trend in our results. The FIP effect modifies only the chromospheric and coronal abundances in the Sun, not the photospheric abundances, which is of relevance here.

\begin{figure}
\centering
\includegraphics[width=\columnwidth]{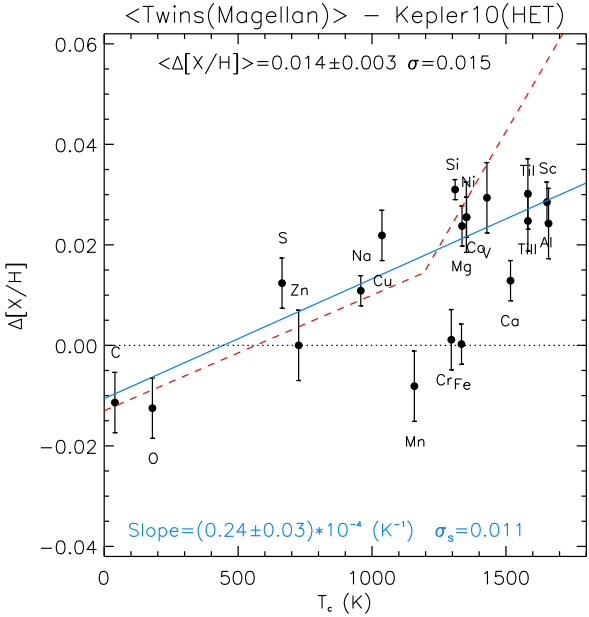}
\caption{Average abundance differences $\Delta$[X/H] versus condensation temperature T$_{\rm c}$ for the eight thick disc stellar twins relative to Kepler-10 with GCE corrections applied. The Y axis is the same as in Figure \ref{fig5}. The blue solid line represents the linear fit to the data, $\sigma_s$ is the dispersion about the linear fit and the red dashed line is the fit from \citet{mel09} for solar twins $-$ Sun, normalized to $\Delta$[C/H].}
\label{fig10}
\end{figure}

Already in discussing Figure \ref{fig3} above we could draw the conclusion that clear correlations with condensation temperature exist for the abundances of Kepler-10 relative to its twins. We have studied this further by applying the linear fits of [X/H] vs. [Fe/H], but using "Twins - Kepler-10" for self-consistency, to correct the abundances of each twin to the [Fe/H] of Kepler-10 and thus derived GCE corrected results. The corrections are relatively small, reflecting the small range in [Fe/H] of the twins. When plotting the differences between these corrected abundances and those of Kepler-10 vs. T$_{\rm c}$ we obtain a diagram (see Figure \ref{fig10}) similar to Figure \ref{fig5}, though with a slightly flatter gradient ((0.24 $\pm$ 0.03) $\times$ 10$^{-4}$ K$^{-1}$) and a marginally larger scatter. Obviously, these GCE corrections can not erase the $\Delta$[X/H] - T$_{\rm c}$ trend.\footnote{A similar approach using Si as the reference element does not change our results. We find a slope of (0.214 $\pm$ 0.036) $\times$ 10$^{-4}$ K$^{-1}$, i.e., a 5.9$\sigma$ result.} One concern regarding the GCE corrections in our analysis is that we are correcting the GCE effects using the abundance ratios - [Fe/H] relations, as was done by \citet{adi14}. \citet{nis15} demonstrated that age may be a better tracer for the Galactic chemical evolution and should be considered when applying the GCE corrections. Indeed a recipe including both age and [Fe/H] could be the best way to estimate the GCE effects. However, the age range of the Kepler-10 thick disc twins is so narrow that we can not, and probably do not need to address an accurate GCE correction using the abundance ratio vs. stellar age plots. Additional thick disc stellar twins would clarify this situation further.

When comparing Kepler-10 to its eight thick disc stellar twins, we find that Kepler-10 is depleted in refractory elements. The $\Delta$[X/H] - T$_{\rm c}$ trends vary star to star which complicates the possible scenario. Chemical signatures of terrestrial planet formation, stellar ages, stellar birth locations, GCE effects, variation in dust-depletion in star-forming regions, etc. may affect our results. We notice that each of the scenarios discussed above may not be fully responsible for the observed abundance while a combination of several factors might affect and produce the current chemical composition of Kepler-10 and its stellar twins.

\section{Conclusions}

We conducted a line-by-line differential abundance study of Kepler-10 and a sample of stellar twins, obtaining extremely high precision based on spectra from three telescopes (CFHT, HET and Magellan). Our analysis reveals subtle chemical differences in the photosphere of Kepler-10 when compared to its stellar twins. We confirm that Kepler-10 is very likely a thick disc star considering its old age (8.4 $\pm$ 1.0 Gyr), kinematic probabilities (96\% as thick disc member) and abundance ratios (according to Figure \ref{fig3}). 
When comparing Kepler-10 to its thick disc twins, a single linear fit provides an appropriate representation of the $\Delta$[X/H] - T$_{\rm c}$ trend. We find that Kepler-10 is depleted in refractory elements relative to volatile elements when compared to the majority of thick disc stellar twins. Two of the eight thick disc twins do not show depletion patterns, which is within the small number statistics compatible with \citet{mel09,ram09,ram10} et al. resulting 15\% of solar twins have chemical compositions that match the solar value. The average abundance difference between thick disc twins and Kepler-10 is 0.037 $\pm$ 0.004 ($\sigma$ = 0.016) which corresponds to at least 13 Earth masses material. One possible explanation could be the formation of terrestrial planets in the Kepler-10 system. However, the results are not as clear as for the solar twins \citep{mel09,ram10}. Other factors (e.g., stellar age, stellar birth location and Galactic chemical evolution) might also affect the abundance results. 

Naturally the thick disc twins may also harbour similarly large rocky planets as Kepler-10 although they have not yet been detected. Several studies based on current discoveries of exoplanets \citep{how12,pet13,bur15} reported estimations of occurrence rate of rocky planets around different type of stars with different orbits. \citet{pet13} indicate that at least one in six stars might host a planet with 1 - 2 R$_{\rm E}$ with period between 5 - 50 days. 
In this case, the peculiar chemical composition of Kepler-10 could reveal signatures regarding the different planetary masses, orbits, formation efficiency or formation timescale. In order to test the \citet{mel09} scenario regarding terrestrial planet formation and unravel the possible subtle chemical signatures and better understand the mechanisms of planet formation, more spectra of terrestrial planets host stars and their identical stellar twins with high SNR ($>$ 350) are needed. It is also important to conduct similar analysis with binary stars (e.g., \citealp{liu14,mac14,tm14,bia15,ram15,saf15,tes15}) or open cluster stars (e.g., \citealp{on11,on14,bru14,spi15}) since these systems presumably share the identical initial chemical composition, thus making them ideal targets for tracing small differential abundance differences that could reveal different formation histories of the individual stars and their planets. 

\section*{Acknowledgments}
This work has been supported by the Australian Research Council (grants FL110100012, FT140100554 and DP120100991). JM thanks support by FAPESP (2012/24392-2). DLL thanks the Robert A. Welch Foundation of Houston, Texas for support through grant F-634. TB was supported by the project grant "The New Milky Way" from the Knut and Alice Wallenberg Foundation.
The Canada-France-Hawaii Telescope is operated by the National Research Council of Canada, the Institut National des Sciences de l'Univers of the Centre National de la Recherche Scientifique of France, and the University of Hawaii. The Hobby-Eberly Telescope is a joint project of the University of Texas at Austin, the Pennsylvania State University, Ludwig-Maximilians-Universit\"{a}t M\"{u}nchen, and Georg-August-Universit\"{a}t G\"{o}ttingen.

\section*{SUPPLEMENTARY MATERIAL}

The following supplementary material is available for this article online:

Table A1. Atomic line data used for our abundance analysis.

Table A2. $\Delta$[X/H] for each program star with relative to Kepler-10. 

\bsp

\label{lastpage}

\end{document}